\documentclass[12pt]{article}
\usepackage{amsmath,amssymb,amsfonts,color,graphicx,cite}
\usepackage{colordvi}
\usepackage{subfigure}
\input paperdef

\graphicspath{{figs/}}

\newcommand{\MHexp}{125.6}

\evensidemargin \oddsidemargin
\allowdisplaybreaks

\begin{document}
\thispagestyle{empty}

\def\thefootnote{\fnsymbol{footnote}}

\begin{flushright}
DESY 14-046\\
FR-PHENO-2014-004\\
MPP--2014--65 
\end{flushright}

\vspace{0.2cm}

\begin{center}

{\Large\sc {\bf Prediction of the light \boldmath{$\cp$}-even 
Higgs-Boson Mass\\[.5em] of the MSSM: Towards the ILC Precision}}%
\footnote{Talk presented by S.H.\ at the International Workshop on
  Future Linear Colliders (LCWS13), Tokyo, Japan, 11-15 November 2013.}

\vspace{0.5cm}

{\sc
T.~Hahn$^{1}$%
\footnote{email: hahn@feynarts.de}%
, S.~Heinemeyer$^{2}$%
\footnote{email: Sven.Heinemeyer@cern.ch}%
, W.~Hollik$^{1}$%
\footnote{email: hollik@mpp.mpg.de}%
, H.~Rzehak$^{3}$%
\footnote{email: hrzehak@mail.cern.ch}%
~and G.~Weiglein$^{4}$%
\footnote{email: Georg.Weiglein@desy.de}
}

\vspace*{.3cm}

{\sl
$^1$Max-Planck-Institut f\"ur Physik (Werner-Heisenberg-Institut),\\
F\"ohringer Ring 6, D--80805 M\"unchen, Germany

\vspace*{0.1cm}

$^2$Instituto de F\'isica de Cantabria (CSIC-UC), Santander, Spain

\vspace*{0.1cm}

$^3$Albert-Ludwigs-Universit\"at Freiburg, 
Physikalisches Institut, D--79104 Freiburg, Germany

\vspace*{0.1cm}

$^4$DESY, Notkestra\ss e 85, D--22607 Hamburg, Germany
}

\end{center}


\begin{abstract}
\noindent
The signal discovered in the Higgs searches at the LHC can be
interpreted as the Higgs boson of the Standard Model as well as the
light $\cp$-even Higgs boson of the Minimal Supersymmetric Standard
Model (MSSM). In this context
the measured mass value, having already reached the level of a precision
observable with an experimental accuracy of about $500 \mev$, plays
an important role. This precision can be improved substantially below
the level of $\sim 50 \mev$ at the future International Linear Collider
(ILC). Within the MSSM the mass of the light
$\cp$-even Higgs boson, $\Mh$, can directly be predicted from
the other parameters of the model. The accuracy of this prediction 
should match the one of the experimental measurements. The
relatively high experimentally observed value of the mass of about
$\MHexp \gev$ 
has led to many investigations where the supersymmetric (SUSY) partners
of the top quark have masses in the multi-TeV range. We review the
recent improvements for the prediction for $\Mh$ in the MSSM
for large scalar top masses.
They were obtained by combining the existing fixed-order
result, comprising the full one-loop and leading and subleading two-loop
corrections, with a resummation of the leading and subleading
logarithmic contributions from the scalar top sector to all orders.
In this way for the first time a high-precision prediction for the mass
of the light $\cp$-even Higgs boson in the MSSM is possible all the way 
up to the multi-TeV region of the relevant supersymmetric particles.
However, substantial further improvements will be needed
to reach the ILC precision. The newly obtained corrections to $\Mh$
are included into the code \fh.

\end{abstract}

\def\thefootnote{\arabic{footnote}}
\setcounter{page}{0}
\setcounter{footnote}{0}

\newpage


\section{Introduction}

After the spectacular discovery of a signal in the Higgs-boson searches
at the LHC by ATLAS and CMS~\cite{ATLASdiscovery,CMSdiscovery}, now
exploration of the properties of the observed particle is in the main
focus. In particular, the observation in the $\ga \ga$ and the
$ZZ^{(*)} \to 4 \ell$ channels made it possible to determine its mass
with already a remarkable precision. Currently, the combined mass
measurement from ATLAS is
$125.5 \pm 0.2 \pm 0.6 \gev$~\cite{ATLAS:2013mma}, and the one
from CMS is $125.7 \pm 0.3 \pm 0.3 \gev$~\cite{CMS:yva}. 
This leads to the naive average of 
\begin{align}
\MH^{\rm LHC, today} &= \MHexp \pm 0.35 \gev~.
\label{MHexp}
\end{align}
At the (planned) future International $e^+e^-$ Linear Collider (ILC),
using the $Z$-recoil method a precision of~\cite{dbd}
\begin{align}
\de\MH^{\rm ILC} &\lsim 50 \mev
\label{Mh-ILCprec}
\end{align}
is currently anticipated.

The other
properties that have been determined so far, in particular the coupling
strength modifiers~\cite{yr3,kappa}, as well as spin, 
are compatible with the minimal realisation of the Higgs
sector within the Standard Model (SM)~\cite{HiggsSMlike}.  
However, a large variety of other
interpretations is possible as well, corresponding  to very different
underlying physics. While within the SM the Higgs-boson mass
is just a free parameter,
in theories beyond the SM (BSM) the mass of the particle that is
identified 
with the signal at about $\MHexp \gev$ can often be directly predicted,
providing an important test of the model. One of the most popular BSM models
is the Minimal Supersymmetric
Standard Model (MSSM)~\cite{mssm}. In this model the Higgs sector
consists of two scalar doublets accommodating five physical Higgs bosons. In
lowest order these are the light and heavy $\cp$-even $h$
and $H$, the $\cp$-odd $A$, and the charged Higgs bosons~$H^\pm$.

The parameters characterising the MSSM Higgs sector at lowest order 
are the gauge couplings, the mass of the $\cp$-odd Higgs boson,
$\MA$, and $\tb \equiv v_2/v_1$,  
the ratio of the two vacuum expectation values. Accordingly, 
all other masses and mixing angles can be predicted in terms of those
parameters, leading to the famous
tree-level upper bound for the mass of the light $\cp$-even Higgs boson, 
$\Mh \le \MZ$, determined by the mass of the $Z$~boson, $\MZ$.
This tree-level upper bound, which arises from the gauge sector,
receives large corrections from the Yukawa sector of the theory, which
can amount up to \order{50\%} (depending on the model
parameters) upon incorporating the full one-loop and the dominant
two-loop contributions~\cite{mhiggsAEC,MHreviews,PomssmRep}.

The prediction for the light $\cp$-even Higgs-boson mass in the MSSM is
affected by two kinds of theoretical uncertainties. First, the
parametric uncertainties induced by
the experimental errors of the input parameters. Here
the dominant source of
parametric uncertainty is the experimental error on the top-quark mass,
$\mt$. Very roughly, the impact of the experimental error on $\mt$ on
the prediction for $\Mh$ scales like~\cite{deltamtILC}
\begin{align}
\de\Mh^{{\rm para,}\mt}/\de\mt^{\rm exp} \sim 1~.
\end{align}
As a consequence, high-precision top-physics providing an accuracy on
$\mt$ much below the GeV-level is a crucial ingredient for precision
physics in the Higgs sector~\cite{deltamtILC}. 
The second type of uncertainties are the intrinsic
theoretical uncertainties that are due to unknown higher-order
corrections. Concerning the SM input parameters, 
an overall estimate of for the lightest $\cp$-even Higgs mass of 
$\de\Mh^{\rm intr} \sim 3 \gev$ 
had been given in \citeres{mhiggsAEC,PomssmRep} (the more recent
inclusion of the leading \order{\alt\als^2} 3-loop
corrections~\cite{mhiggsFD3l}, see below, has slightly reduced this estimated
uncertainty by few times \order{100 \mev}). It was pointed out
that a more detailed estimate needs to take into account the dependence
on the considered parameter region of the model. In particular, the
uncertainty of this fixed-order prediction is somewhat larger 
for scalar top masses in the multi-TeV range.

The MSSM parameter space with scalar top masses in the multi-TeV range
has received considerable attention recently, partly 
because of the relatively high value of $\Mh \approx \MHexp \gev$, which
generically requires either large stop masses or large mixing in the
scalar top sector, and partly because of the limits from searches for 
supersymmetric (SUSY) particles at the LHC.
While within the general MSSM the lighter scalar 
superpartner of the top quark is allowed to be relatively light (down to
values even as low as $\mt$), both with respect to the direct searches 
and with respect to the prediction for $\Mh$ (see
e.g.\ \citere{Mh125}), the situation is different in more constrained
models. For instance, global fits in the Constrained MSSM (CMSSM) prefer
scalar top masses in the multi-TeV range~\cite{mc8,mc9,fittino}. 

Here we review the significantly improved prediction for the
mass of the light $\cp$-even Higgs boson in the MSSM~\cite{Mh-logresum},
which is expected to have an important impact on the phenomenology in the
region of large squark masses and on its confrontation 
with the experimental results. We briefly review the relevant sectors
and the new, improved prediction for $\Mh$. The numerical analysis
focuses on the effects of heavy scalar top masses. The feasability of
reaching the anticipated ILC precision will be briefly discussed.


\section{The Higgs and scalar top sectors of the MSSM}

In the MSSM with real parameters (we restrict to this case for
simplicity; for the treatment of complex parameters see
 \citeres{mhcMSSMlong,mhcMSSM2L} and references therein),
using the Feynman diagrammatic (FD) approach,
 the higher-order corrected  
$\cp$-even Higgs-boson masses are derived by finding the
poles of the $(h,H)$-propagator 
matrix. The inverse of this matrix is given by
\begin{align}
-i \; 
\ML  p^2 -  \mhtree^2 + \hSi_{hh}(p^2)  &  \hSi_{hH}(p^2) \\
     \hSi_{hH}(p^2) & p^2 -  \mHtree^2 + \hSi_{HH}(p^2) \MR,
\label{higgsmassmatrixnondiag}
\end{align}
where $m_{h,H, {\rm tree}}$ denote the 
tree-level masses, 
and $\hSi_{hh,HH,hH}(p^2)$ are the renormalized Higgs boson
self-energies evaluated at the squared external momentum~$p^2$.
For the computation of the leading contributions to those
self-energies it is convenient to use the basis of the fields $\phi_1$,
$\phi_2$, which are related to $h$, $H$ via the (tree-level)
mixing angle $\al$: 
\begin{align}
\ML h \\ H \MR &= \ML -\Sa & \Ca \\ \Ca & \Sa \MR \ML \phi_1 \\ \phi_2 \MR~.
\end{align}

\bigskip
The new higher-order corrections reviewed here originate in the top/stop
sector of the MSSM. 
The bilinear part of the top-squark Lagrangian,
\begin{align}
\cL_{\Stop, \text{mass}} &= - \begin{pmatrix}
{{\tilde{t}}_{L}}^{\dagger}, {{\tilde{t}}_{R}}^{\dagger} \end{pmatrix}
\matr{M}_{\tilde{t}}\begin{pmatrix}{\tilde{t}}_{L}\\{\tilde{t}}_{R}
\end{pmatrix} ~,
\end{align}
contains the stop mass matrix, $\matr{M}_{\tilde{t}}$,
given by 
\begin{align}\label{Sfermionmassenmatrix}
\matr{M}_{\tilde{t}} &= \begin{pmatrix}  
 \MstL^2 + \mt^2 + \MZ^2 \CZb \, (T_t^3 - Q_t \sw^2) & 
 \mt \Xt \\[.2em]
 \mt \Xt &
 \MstR^2 + \mt^2 + \MZ^2 \CZb \, Q_t \, \sw^2
\end{pmatrix}~, \\
{\rm with} &\mbox{} \non \\
\Xt &= \At - \mu\,\CTb~.
\end{align}
$Q_t$ and $T_t^3$ denote charge and isospin of the top quark, 
$\At$ is the trilinear coupling between the Higgs bosons and the scalar
top quarks, and $\mu$ is the Higgs mixing parameter.
The mass matrix can be diagonalized with the help of a unitary
transformation ${\matr{U}}_{\tilde{t}}$, yielding the two stop
mass eigenvalues, $\mste$ and $\mstz$.

\bigskip
For the MSSM with real parameters
the status of higher-order corrections to the masses and mixing angles
in the neutral Higgs sector is quite advanced. The complete one-loop
result within the MSSM is known~\cite{ERZ,mhiggsf1lA,mhiggsf1lB,mhiggsf1lC}.
The by far dominant one-loop contribution is the \order{\alt} term due
to top and stop loops ($\alt \equiv h_t^2 / (4 \pi)$, $h_t$ being the
top-quark Yukawa coupling). The computation of the two-loop corrections
has meanwhile reached a stage where all the presumably dominant
contributions are 
available~\cite{mhiggsletter,mhiggslong,mhiggslle,mhiggsFD2,bse,mhiggsEP0,mhiggsEP1,mhiggsEP1b,mhiggsEP2,mhiggsEP3,mhiggsEP3b,mhiggsEP4,mhiggsEP4b,mhiggsRG1,mhiggsRG1a}.
In particular, the \order{\alt\als} contributions to the self-energies
-- evaluated in the 
Feynman-diagrammatic (FD) as well as in the effective potential (EP)
method -- as well as the \order{\alt^2}, \order{\alb\als}, 
\order{\alt\alb} and \order{\alb^2} contributions  -- evaluated in the EP
approach -- are known. (For latest corrections to the charged Higgs
boson mass, see \citere{chargedHiggs2L}.) 

The public code
\fh~\cite{feynhiggs,mhiggslong,mhiggsAEC,mhcMSSMlong,Mh-logresum}  
includes all of the above corrections, where the on-shell (OS) scheme for the
renormalization of the scalar quark sector has been used (another
public code, based on the Renormalization Group (RG) improved
Effective Potential, is {\tt CPsuperH}~\cite{cpsh}).
A full 2-loop effective potential calculation %
(supplemented by the momentum dependence for the leading
pieces and the leading 3-loop corrections) has been
published~\cite{mhiggsEP5,mhiggsRGE3l}. However, no computer code 
is publicly available. Most recently another leading 3-loop
calculation at \order{\alt\als^2} became available (based on a
\DRbar\ or a ``hybrid'' renormalisation scheme for the scalar top
sector), where the numerical
evaluation depends on the various SUSY mass hierarchies~\cite{mhiggsFD3l}, 
resulting in the code {\tt H3m} (which
adds the 3-loop corrections to the \fh\ result).


\section{Improved calculation of \boldmath{$\Mh$}}

We review here the improved prediction for $\Mh$ where we combine the
fixed-order result obtained in the OS scheme with an all-order
resummation of the leading and subleading contributions from the scalar
top sector. We have obtained the latter from an analysis of the 
RG Equations (RGEs) at the two-loop level~\cite{SM2LRGE}.
Assuming a common mass scale $\MS = \sqrt{\mste\,\mstz}$ ($\MS \gg \MZ$)
for all relevant SUSY mass parameters,
the quartic Higgs coupling $\la$ can be evolved
via SM RGEs from \MS\ to the scale $Q$ 
(we choose $Q = \mt$ in the following)
where $\Mh^2$ is to be evaluated
(see, for instance, \citere{bse} and references therein),
\begin{align}
\Mh^2 = 2 \la(\mt) v^2~.
\label{Mh2RGE}
\end{align}
Here $v \sim 174 \gev$ denotes the vacuum expectation value of the SM. 
Three coupled RGEs, the ones for
\begin{align}
\la, \; h_t, \; g_s
\end{align} 
 are relevant for this evolution, 
with the strong coupling constant given as $\als = g_s^2/(4\,\pi)$. 
Since SM RGEs are used, the relevant parameters are given in the
\MSbar\ scheme.
We incorporate the one-loop threshold corrections to
$\la(\MS)$ as given in \citere{bse},
\begin{align}
\la(\MS) = \frac{3\,h_t^4(\MS)}{8\,\pi^2} \frac{\Xt^2}{\MS^2} 
           \KKL 1 - \ed{12} \, \frac{\Xt^2}{\MS^2} \KKR~, 
\label{threshold}
\end{align}
where as mentioned above $\Xt$ is an \MSbar\ parameter. 
Furthermore, in \refeq{threshold} we have set the SM gauge couplings to 
$g = g' = 0$, ensuring that \refeq{Mh2RGE} consists of the ``pure loop
  correction'' and will be denoted $(\De\Mh^2)^{\rm RGE}$ below.
Using RGEs at two-loop order~\cite{SM2LRGE}, 
including fermionic contributions from the top sector only, 
leads to a prediction for the corrections to $\Mh^2$ including 
leading and subleading logarithmic contributions at $n$-loop order, 
\begin{align}
L^n \mbox{~and~} L^{(n-1)}, \; L \equiv  \ln\KL \frac{\MS}{\mt} \KR~,
\end{align}
originating from the top/stop sector of the MSSM.

We have obtained both analytic solutions of the RGEs up to the $7$-loop
level as well as a numerical solution incorporating the leading and
subleading logarithmic contributions up to all orders.
In a similar way in \citere{mhiggsRGE3l} the leading logarithms at
3- and 4-loop order have been evaluated analytically.
Most recently a calculation using 3-loop SM RGEs
appeared in \citere{Carlos-3L-RGE}.

\smallskip
A particular complication arises in 
the combination of the higher-order logarithmic contributions
obtained from solving the RGEs with the 
fixed-order FD result implemented in \fh\ comprising corrections up to the
two-loop level in the OS scheme. We have used the parametrisation of the
FD result in terms of the running top-quark mass at the scale $\mt$, 
\begin{align}
\overline{\mt} &=
\frac{\mt^{\rm pole}}{1 + \frac{4}{3 \pi} \als(\mt^{\rm pole}) 
-\ed{2 \pi} \alt(\mt^{\rm pole})}~,
\end{align} 
where $\mt^{\rm pole}$ denotes the top-quark pole mass. 
Avoiding double counting of the logarithmic contributions up to the
two-loop level and consistently taking into account the different
schemes employed in the FD and the RGE approach, the correction
$\De\Mh^2$ takes the form
\begin{align}
\De\Mh^2 &= (\De\Mh^2)^{\rm RGE}(\Xt^{\MSbar}) 
          - (\De\Mh^2)^{\rm FD, LL1,LL2}(\Xt^{\OS})~, \non \\[.3em]
\Mh^2 &= (\Mh^2)^{\rm FD} + \De\Mh^2~.
\label{eq:combcorr}
\end{align} 
Here $(\Mh^2)^{\rm FD}$ denotes the fixed-order FD result,
$(\De\Mh^2)^{\rm FD, LL1, LL2}$ are the logarithmic contributions up to
the two-loop level obtained with the FD approach in the OS scheme, while 
$(\De\Mh^2)^{\rm RGE}$ are the leading and sub-leading logarithmic
contributions (either up to a certain loop order or summed to all
orders) obtained in the RGE approach, as evaluated via \refeq{Mh2RGE}. 
In all terms of
\refeq{eq:combcorr} the top-quark mass is parametrised in terms of
$\overline{\mt}$;
the relation between $\Xt^{\MSbar}$ and $\Xt^{\OS}$ is given by
\begin{align}
\Xt^{\MSbar} = \Xt^{\OS} \KKL 1 + 
2 L \KL \frac{\als}{\pi} - \frac{3\, \alt}{16\,\pi} \KR \KKR
\end{align}
up to non-logarithmic terms, and there are no logarithmic contributions 
in the relation between $\MS^{\MSbar}$ and $\MS^{\OS}$. 

Since our higher-order contributions beyond 2-loop
have been derived under the assumption
$\MA \gg \MZ$, to a good approximation these corrections
can be incorporated 
as a shift in the prediction for the $\phi_2\phi_2$ self-energy 
(where $\De\Mh^2$ 
enters with a coefficient $1/\SQb$). In this way
the new higher-order contributions enter not only the prediction for
$\Mh$, but also consistently the ones all other Higgs sector observables
that are evaluated in  \fh, such as the effective mixing angle $\aeff$ or the
finite field renormalization constant matrix
$\matr{Z_n}$~\cite{mhcMSSMlong}. 

\medskip
The latest version of the code, \fh\,{\tt 2.10.0}, which is
available at {\tt feynhiggs.de}, contains those improved
predictions as well as a refined estimate of the theoretical
uncertainties from unknown higher-order corrections. Taking into account
the leading and subleading logarithmic contributions in higher orders
reduces the uncertainty of the remaining unknown higher-order
corrections. Accordingly, the estimate of the uncertainties arising from
corrections beyond two-loop order in the top/stop sector is adjusted
such that the impact of replacing the running top-quark mass by the pole
mass (see \citere{mhiggsAEC}) is evaluated only for the non-logarithmic
corrections rather than for the full two-loop contributions implemented
in \fh. First investigations using this new uncertainty estimate can be
found in \citeres{ehowp,mc9}.

\medskip
Further refinements of the RGE resummed result are possible, in
particular 
\begin{itemize}
\item
extending the result to the case of a large
splitting between the left- and right-handed soft
SUSY-breaking terms in the scalar top sector~\cite{mhiggsEP3b},
\item
extending the result to the case of small $\MA$ or $\mu$ (close to
$\MZ$), 
\item
including the corresponding contributions from the (s)bottom sector. 
\end{itemize}
Some details in these directions can be found in \citere{Carlos-3L-RGE}.
We leave those refinements for future work.


\section{Numerical analysis}

\begin{figure}[htb!]
\centerline{
\includegraphics[width=0.65\textwidth,height=09.0cm]{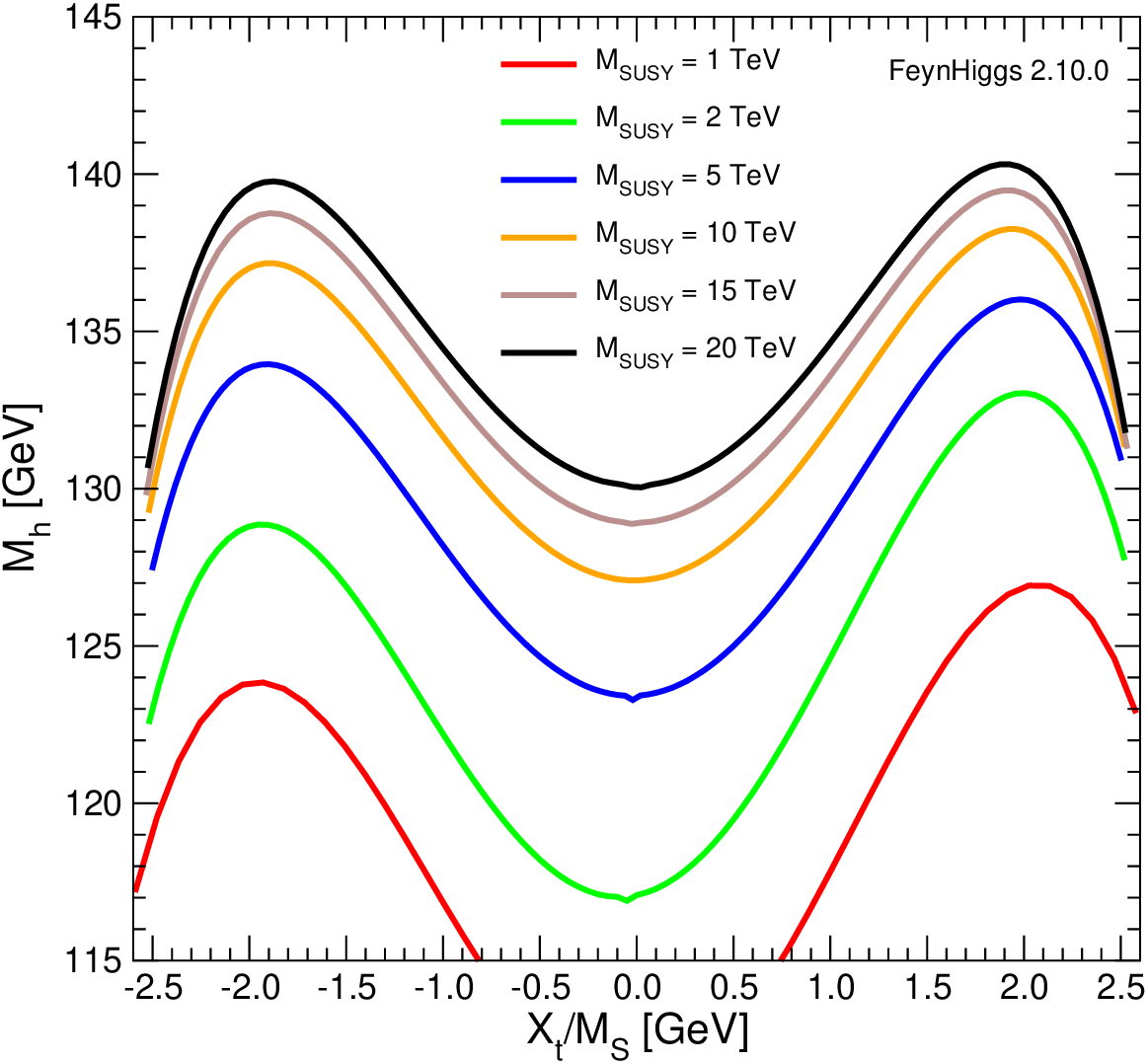}
}
\caption{
$\Mh$ as a function of $\Xt/\MS$ for various values of $\MS$, 
obtained using the full result as implemented into \fh\,{\tt 2.10.0}.
}
\label{fig:plot1}
\end{figure}

\begin{figure}[htb!]
\vspace{1em}
\centerline{
\includegraphics[width=0.65\textwidth,height=09.0cm]{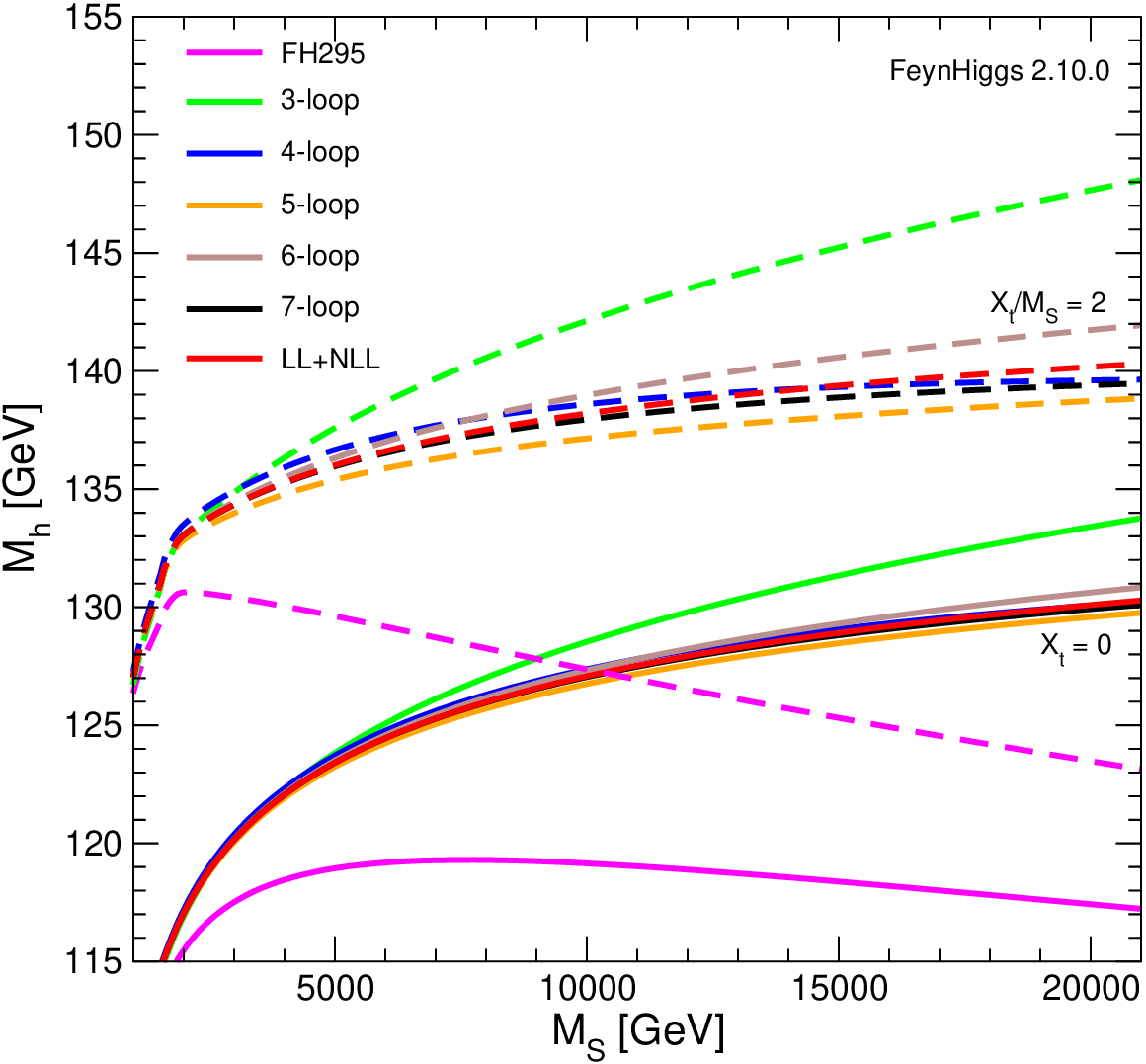}
}
\caption{
$\Mh$ as a function of $\MS$ for $\Xt = 0$ (solid) and 
$\Xt/\MS = 2$ (dashed). The full result (``LL+NLL'') is compared with
results containing the logarithmic contributions up to the 3-loop,
\ldots 7-loop level and with
the fixed-order FD result (``FH295'').
}
\label{fig:plot2}
\vspace{-1em}
\end{figure}

\begin{figure}[htb!]
\centerline{
\includegraphics[width=0.65\textwidth]{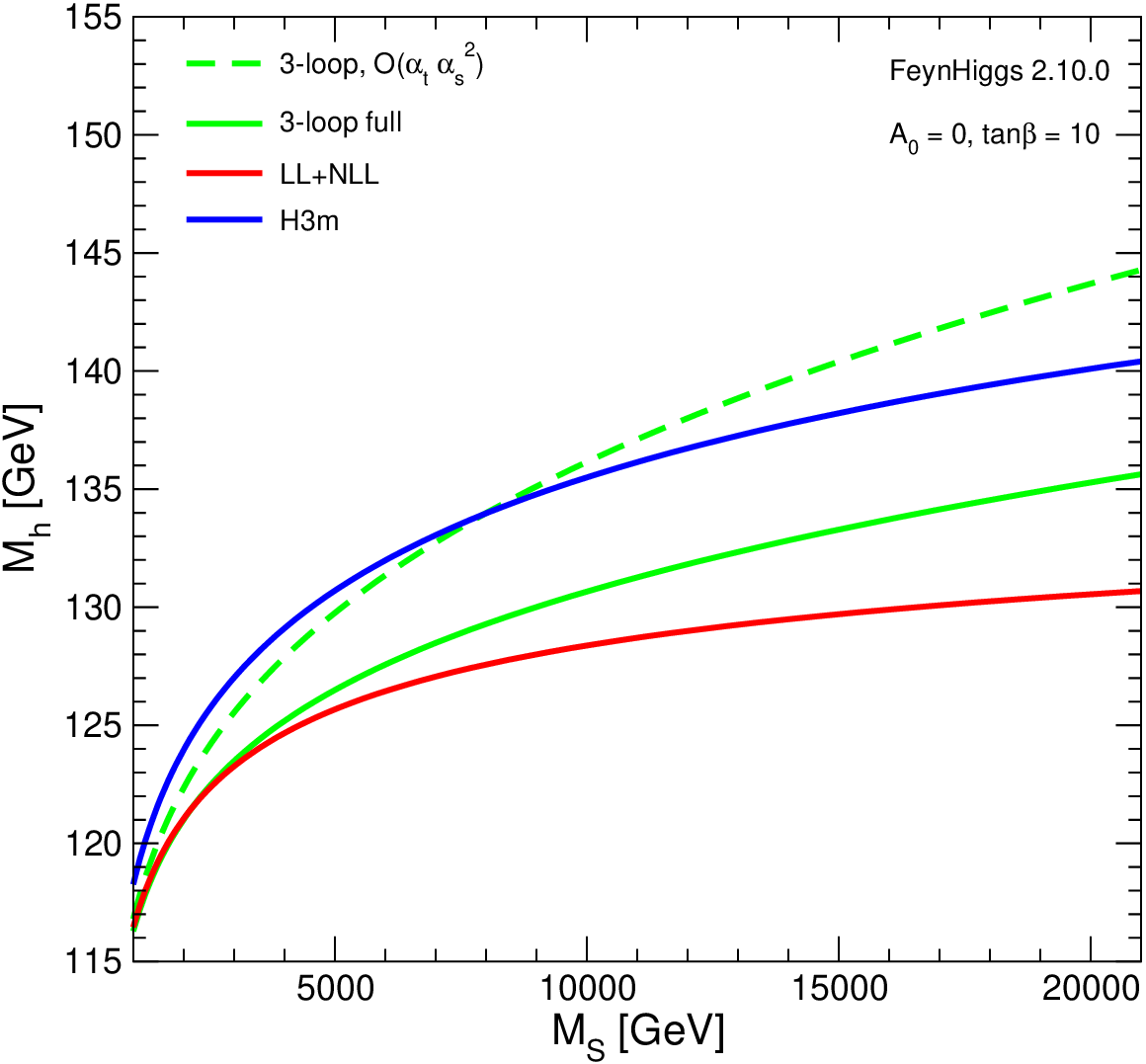}
}
\caption{
$\Mh$ as a function of $\MS$.
Comparison of \fh\ (red) with {\tt H3m} (blue). 
In green we show the \fh\ 3-loop result at \order{\alt\als^2} 
(\order{\alt\als^2, \alt^2\als, \alt^3}) as
dashed (solid) line.
}
\label{fig:plot3}
\end{figure}

In this section we review the analysis of the phenomenological implications of
the improved $\Mh$ prediction for large stop mass scales, as
evaluated with \fh\,{\tt 2.10.0}.
Here and in the following $\Xt$ denotes $\Xt^{\OS}$ (for $\MS$ the
difference in the two schemes is negligible).
The other parameters are $\MA = M_2 = \mu = 1000 \gev$,
$\mgl = 1600 \gev$ (where $M_2$ is the SU(2) gaugino mass term, $\mu$ the
Higgsino mass parameter and $\mgl$ the gluino mass) and $\tb = 10$.

In \reffi{fig:plot1} we show $\Mh(\Xt/\MS)$ for various values of $\MS$
(as indicated by different colors), evaluated with the full new result
as implemented into \fh\,{\tt 2.10.0}. 
It can be seen that local maxima are reached for $\Xt/\MS \approx \pm 2$,
where the $\Mh$ values at $\Xt/\MS \approx +2$ are slightly larger than
the ones at $\Xt/\MS \approx -2$. 
Local minima are reached around $\Xt/\MS \approx 0$. This
feature was well known for the results at the 2-loop level, and now are
shown to persist for the inclusion of the resummed leading and
subleading logarithmic corrections. 
It can furthermore be seen that the current experimental value of 
$\Mh = (\MHexp \pm 0.35) \gev$ can be reached for many combinations of
$\Xt$ and $\MS$ (for the other parameters fixed as given above), but on
the other hand, that many of such combinations can also be ruled out by
the Higgs-boson mass constraint. For a more detailed analysis the
theoretical uncertainties have to be taken into account properly, see
the discussion below. 

The results of \reffi{fig:plot1} motivate the choice of parameters used
in \reffi{fig:plot2}. Here we show $\Mh$ as a function of 
$\MS$ for $\Xt = 0$ and $\Xt/\MS = +2$, corresponding to the local minimum
and the maximum value of $\Mh$ as a function of $\Xt/\MS$, respectively.
The plot shows for the two values of $\Xt/\MS$ the
fixed-order FD result containing corrections up to the two-loop level
(labelled as ``FH295'', which refers to the previous version of the code
\fh) as well as the latter result supplemented with the analytic
solution of the RGEs up to the 3-loop, \ldots 7-loop level 
(labelled as ``3-loop'' \ldots ``7-loop''). The curve labelled as
``LL+NLL'' represents our full result, where the FD contribution is
supplemented by the leading and next-to-leading logarithms summed to all
orders.
One can see that the impact of the higher-order logarithmic
contributions is relatively small for $\MS = \order{1 \tev}$, while
large differences between the fixed-order result and the improved
results occur for large values of $\MS$.
The 3-loop logarithmic
contribution is found to have the largest impact in this context, but 
for $\MS \gsim 2500 (6000) \gev$ for
$\Xt/\MS = 2 (0)$ also contributions beyond 3-loop are important.
A convergence of the higher-order logarithmic contributions towards the
full resummed result is clearly visible.
At $\MS = 20 \tev$ the 
difference between the 7-loop result and the full resummed result is
around $900 (200) \mev$ for $\Xt/\MS = 2 (0)$. The corresponding
deviations stay below $100 \mev$ for $\MS \lsim 10 \tev$.
The plot furthermore shows that for $\MS \approx 7 \tev$ (and the value
of $\tb = 10$ chosen here) a predicted value of $\Mh$ of about $\MHexp \gev$
is obtained even for the case of vanishing mixing in the scalar
top sector ($\Xt = 0$). Since the predicted value of $\Mh$ grows further
with increasing $\MS$ it becomes apparent that the measured 
mass of the observed signal, when interpreted as $\Mh$, can
be used (within the current experimental
and theoretical uncertaintes) to derive an {\em upper bound\/} on the
mass scale $\MS$ in the scalar top sector,
see also \citere{StopUpperLimit}. However, more robust statements in
this direction will require a careful analysis of still present
intrinsic as well as the parametric uncertainties.

Finally, in \reffi{fig:plot3} we compare our result with the
prediction obtained from the code {\tt H3m}~\cite{mhiggsFD3l}.
The comparison was performed in the
CMSSM with the parameters set to 
$m_0 = m_{1/2} = 200 \gev \ldots 15000 \gev$, 
$A_0 = 0$, $\tb = 10$ and $\mu > 0$. The spectra were generated with 
{\tt SoftSusy\, 3.3.10}~\cite{softsusy}.
The {\tt H3m} result shown as blue line, containing the terms in 
\order{\alt\als^2}$\times$\order{L^3, L^2, L^1, L^0}, can be compared
with the \fh\ 3-loop result, \order{L^3, L^2}, but restricted to
\order{\alt\als^2} (green dashed).
We find that the latter result agrees rather well with {\tt H3m},
with maximal deviation of \order{1 \gev} for 
$\MS \lsim 10 \tev$. The observed deviations 
can be attributed to the terms of \order{L^1, L^0} included in 
{\tt H3m}, to the SUSY mass hierarchies taken into account in {\tt H3m},
and to the use of different scalar top renormalization schemes employed
in the two codes (where the latter effect is already expected to be at the
GeV-level). Further investigations will be needed to explore the 
source of the main differences.
However, adding also the
3-loop  \order{\alt^2\als, \alt^3}$\times$\order{L^3, L^2}
terms (solid green), as included in the \fh\ result, leads to a strong
reduction of $\Mh$ by $\sim 5 \gev$ for $\MS = 10 \tev$ (see also
\citere{mhiggsRGE3l}). Going to the
full resummed \fh\ result (red) exhibits a further, but smaller
reduction of $\Mh$ of about $2 \gev$ for $\MS = 10 \tev$, 
even larger changes are found for $\MS > 10 \tev$. 
Consequently, 3-loop corrections at
\order{\alt^2\als, \alt^3} as well as corrections beyond
3-loop are clearly important for a precise $\Mh$ prediction.  

\smallskip
In view of the anticipated future accuracy at the ILC, as given in
\refeq{Mh-ILCprec}, the remaining theory uncertainties in the current
status of the $\Mh$ evaluations will have to be re-analyzed carefully. It can
be expected, see also \citere{ehowp}, that for scalar top mass scales
below the few-TeV level the intrinsic uncertainty is now,
i.e.\ including the resummed contributions, at or below the level of
$\sim 2 \gev$. However, still substantial further refinements will be
needed to reach the sub-GeV level. On the other hand, no investigation
of the size of the intrinsic uncertainties has been performed for scalar
top masses in the multi-TeV range, as explored here. Consequently, the
prospects of reaching the sub-GeV level in $\de\Mh^{\rm intr}$ are so
far unclear. Turning to the parametric uncertainties, the ILC precision
for $\mt$ with $\de\mt^{\rm ILC} \lsim 100 \mev$~\cite{dbd} will be
crucial to lower the $\mt$-induced uncertainty to the level of the
anticipated ILC 
accuracy for $\Mh$. Being stuck with a hadron collider precision on
$\mt$ will always induce a parametric uncertainty in the $\Mh$
prediction substantially larger than even the current experimental
uncertainty as given in \refeq{MHexp}.


\section{Conclusions}

We have reviewed the improved prediction for the light
$\cp$-even Higgs-boson mass in the MSSM, obtained by combining the
FD result at the one- and two-loop level with an
all-order resummation of the leading and subleading logarithmic 
contributions from the top/stop sector obtained from solving the two-loop
RGEs. Particular care has been taken
to consistently match these two different types of corrections. The result,
providing the most precise prediction for $\Mh$ in the presence of large
masses of the scalar partners of the top quark, has been implemented
into the public code \fh\ and can be obtained at 
{\tt feynhiggs.de}\,. We have found a sizable effect of the
higher-order logarithmic contributions for 
$\MS \equiv \sqrt{\mste\mstz} \gsim 2 \tev$ which grows with increasing
$\MS$. 
In comparison with {\tt H3m}, which calculates the
  \order{\alt\als^2} corrections to $\Mh$ we find that both,
corrections
 at 3-loop at \order{\alt^2\als,\alt^3} as well as corrections beyond
3-loop  are important for a precise $\Mh$ prediction; 
for $\MS = 10 \tev$
differences of $\sim 7 \gev$ are found between {\tt H3m} and \fh\ (for
the parameters used in our numerical analysis).
Finally, we have shown that for sufficiently high $\MS$ the predicted
values of $\Mh$ reaches about $\MHexp \gev$ even for vanishing mixing in the
scalar top sector. As a consequence, even higher $\MS$ values are
disfavoured by the measured mass value of the Higgs signal.

Reaching an intrinsic uncertainty in the $\Mh$ prediction at the sub-GeV
level will require the inclusion of substantially more higher-order
corrections. This accuracy will be crucial even to meet the current LHC
precision. Reaching the future LHC precision or even the ILC precision
of $\Mh$ will be correspondingly more demanding.


\subsection*{Acknowledgements}

We thank 
H.~Haber, 
P.~Kant, 
P.~Slavich 
and 
C.~Wagner
for helpful discussions.
The work of S.H.\ was supported by the 
Spanish MICINN's Consolider-Ingenio 2010 Program under Grant MultiDark No.\ 
CSD2009-00064. 
The work of G.W.\ was supported by the Collaborative
Research Center SFB676 of the DFG, ``Particles, Strings, and the early
Universe".



\end{document}